\documentclass[aps,prb,amsmath,twocolumn,amssymb,notitlepage,superscriptaddress]{revtex4-1}	
\usepackage[english]{babel}
\usepackage{graphicx}
\usepackage{amsmath}
\usepackage{amssymb}
\usepackage{epstopdf}
\usepackage{amsfonts}
\usepackage{bm}
\usepackage{mathdots}
\usepackage{times}
\usepackage{mathtools}
\usepackage{hyperref}
\usepackage{amsmath}
\usepackage{amssymb}
\usepackage{graphicx}
\usepackage{tikz}

\usepackage{color}

\newcommand{\astcycl}{\mathrlap{\kern0.085em{\circlearrowright}}\ast}
\newcommand{\taucycl}{\mathrlap{\kern0.42em{\bullet}}\circlearrowright}

\newcommand{\mathbbm}[1]{{\bm #1}}

\begin{document}
\title{Collective excitations of the U(1)-symmetric exciton insulator in a cavity}

\author{Katharina Lenk}
\affiliation{Department of Physics, University of Erlangen-N\"urnberg, 91058 Erlangen, Germany}
\author{Martin Eckstein}
\affiliation{Department of Physics, University of Erlangen-N\"urnberg, 91058 Erlangen, Germany}

\begin{abstract}
We investigate the equilibrium state and the collective modes of an excitonic insulator (EI) in a 
Fabry-P\'erot 
cavity. In an EI, two bands of a semiconductor or semimetal spontaneously hybridize due to the Coulomb interaction between electrons and holes, leading to the opening of a gap. The coupling to the electromagnetic field reduces the symmetry of the system with respect to phase rotations of the excitonic order parameter from $U(1)$ to $Z_2$. While the reduction to a discrete symmetry would in general 
lead to a gapped phase mode and enhance the stability of the ordered phase, the  coupling to the  cavity
leaves the mean-field ground state unaffected. Its energy remains invariant under $U(1)$ phase rotations, in 
spite of the lower $Z_2$ symmetry imposed by the cavity.  In dipolar gauge, this can be traced back to the balancing of the linear light-matter coupling and the dipolar self-interaction at zero frequency. At nonzero frequency, however, the collective excitations do reflect the  lower $Z_2$ 
symmetry, which shows that fluctuations beyond mean-field could play a crucial role in finding the true phase at finite temperature.
\end{abstract}
\date{\today}

\maketitle

\section{Introduction}

Novel quantum states can arise when matter is driven by electromagnetic fields beyond linear response. While many interesting examples of classical light-driven dynamics in solids have been reported,\cite{Basov2017} yet a new class of unexplored phenomena is encountered when the quantum nature of the electromagnetic field becomes relevant.  This is achieved by shaping the field in a cavity, so that the light-matter coupling is enhanced  to the point where a single photon becomes relevant.  The advance of cavity quantum electrodynamics (QED) into this ultra-strong coupling regime \cite{Frisk-Kockum2019} has inspired many proposals for hybrid light-matter states in solids.\cite{Schachenmayer2015, Kiffner2019, Schlawin2019, Sentef2018b, Mazza2019, Wang2019, Curtis2019, Kiffner2019b, Orgiu2015}  
An interesting situation arises when the coupling to the field changes the symmetry of a system,  and thus alters the nature of its symmetry-broken states. For example, one can consider a system with a complex order parameter $\Phi$ which breaks a continuous $U(1)$ symmetry. 
If  the  cavity mode $Q\in\mathbb{R}$ and the order parameter are linearly coupled, $F(Q,\Phi) = a|\Phi|^2+b|\Phi|^4 + \gamma Q(\Phi+\Phi^*) + \omega Q^2$, the symmetry is reduced to from $U(1)$ to $Z_2$. This should 
add a mass to the phase mode and  enhance the stability of the symmetry-broken phase. We will investigate this situation for 
the ferroelectric excitonic insulator (EI), 
a quantum phase which has achieved considerable attention  in the context of classical light-driven phenomena. The transition to the EI is driven by the Coulomb interaction, and its order parameter involves a spontaneous hybridization $\langle c^\dagger f\rangle$ between two bands $c$ and $f$ of a semimetal or insulator.\cite{Mott1961,RevModPhys.40.755, Jerome1967}  Materials which are supposed to host an EI phase are Ta${}_{2}$NiSe${}_{5}$  \cite{Kaneko2013,mazza2019b} and 1T-TiSe$_2$, where the softening of the exciton  mode at the transition has been observed recently.\cite{Kogar2017} The electronic nature of the phase transition makes the EI of interest to study the light-induced dynamics. Several experimental and theoretical studies have focused on the photo-induced melting  \cite{Rohwer2011,Mathias2016,Golez2016} and the characterization of the collective modes,\cite{Porer2014,Werdehausen2018,Kogar2017} and it has been shown that the gap in Ta${}_{2}$NiSe${}_{5}$  can be enhanced through photo-doping.\cite{mor2017} 

If the charge in the individual bands is conserved, the EI breaks a $U(1)$ symmetry.  This implies a softening of the collective exciton mode in the normal state towards the transition, and a massless phase mode in the symmetry-broken phase.  
The effect of coupling a real mode $Q$ 
to the $U(1)$-symmetric EI has been studied in the framework of an electron-phonon coupling,\cite{Zenker2014, Zenker2013, Kaneko2013, Murakami2017} where a massive phase mode is observed.  
The same is found when the EI is linearly coupled to the $q=0$ mode of the transverse vector potential.\cite{Mazza2019}
If the corresponding inter-band transition carries a dipole moment, the EI becomes an electronically driven ferroelectric.\cite{Portengen1996, Batista2002, Zenker2010b} The phase then linearly couples to the electric field of a cavity mode $Q$, and one could expect a similar effect on the EI as for the phonon. However,  the mean-field state is not affected by the cavity. In dipolar gauge, this is explained by a balancing of the linear $D\cdot P$ coupling of the displacement field  to the polarization, and the dipolar self-interaction $P\cdot P$. As a result, the mean-field state still has an arbitrary $U(1)$ phase, and a massless phase mode, in agreement with a similar study by Andolino et al.\cite{Andolina2019} At nonzero frequencies, however, the  effect of the two interactions on the EI does not cancel, so that the collective properties at $\omega >0$ do depend on the phase of  the order parameter. This result could be important  to understand pathways  for controlling the EI phase in a cavity,  and it also  highlights  the crucial and often subtle choice of the correct 
light-matter Hamiltonian at strong coupling, which has been discussed widely in the context of cavity QED.\cite{Di-Stefano2019,Bernardis2018b,Li2020} 

The paper is outlined as follows. In Sec.~\ref{sec:model} we define the model for the two-band EI in a cavity in dipolar gauge, and provide a detailed outline of the mean-field solution. Section \ref{sec:results} presents the results for the collective modes of the system, and Section~\ref{sec:conclusion} contains a conclusion and discussion.

\section{Model and methods}
\label{sec:model}

\subsection{Dipolar light-matter Hamiltonian}
\label{sec:lmham}

\begin{figure}
	\centerline{
	\includegraphics[width=0.5\textwidth]{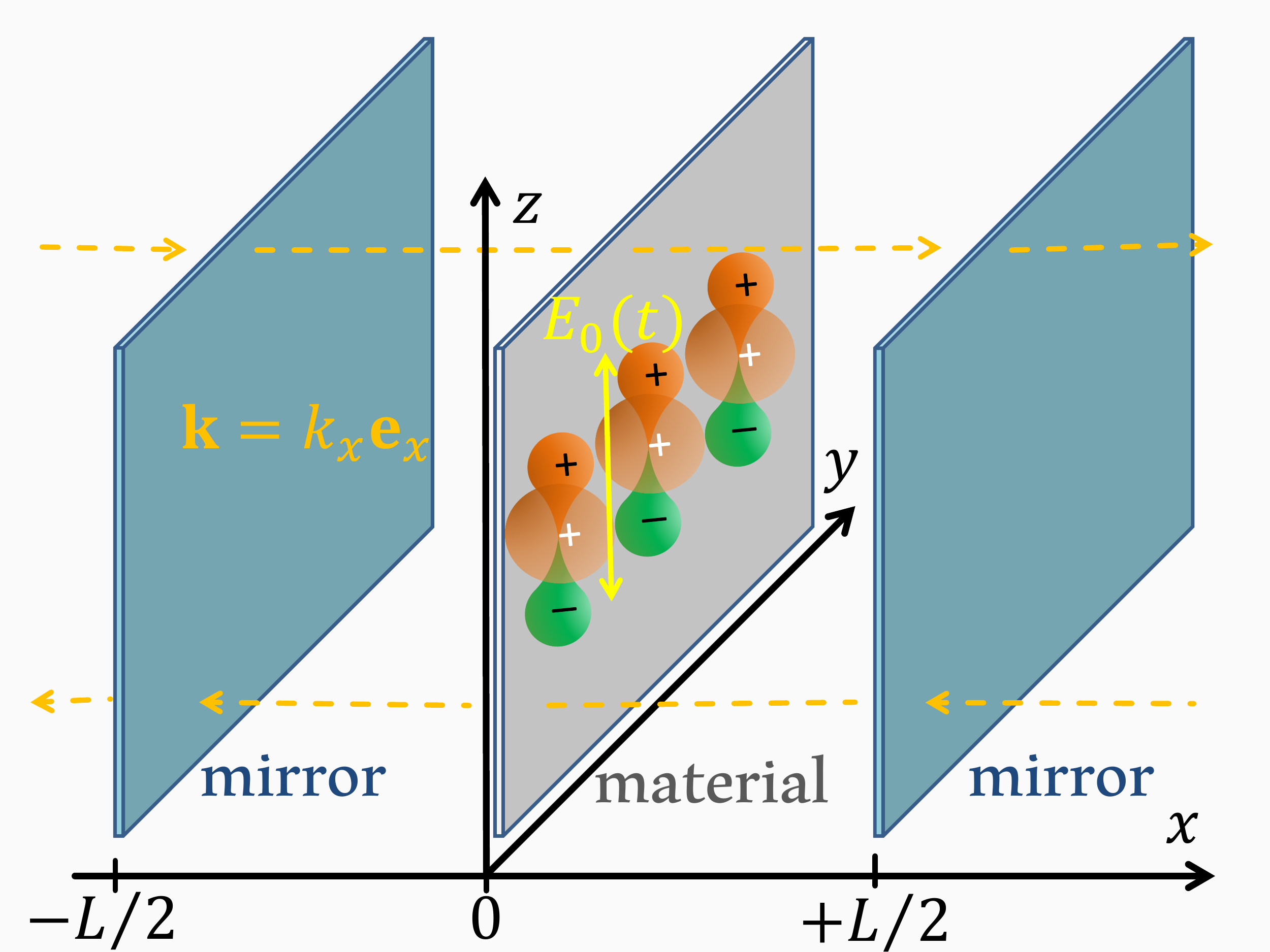}}
	\caption{Setup: The system consists of two parallel plane mirrors at $x=\pm L/2$ and a thin slab of material at $x=0$. It contains two electronic bands with orbitals of opposite parity at each lattice site (e.g., $s$-like and $p_z$-like orbitals as shown in the graphic). Therefore, a hybridization of the bands leads to a finite polarization of the material. For simplicity, we only take into account modes of the electric field that are travelling in $x$-direction, and it it is assumed that the orientation of the orbitals only allows the $z$-component of the electromagnetic field to couple to the material. Within the $y$-$z$-plane the entire system is translation invariant and the electromagnetic field is homogenous.}
	\label{fig:setup}
\end{figure}

We consider a geometry as sketched in Fig.~\ref{fig:setup}, with a 
Fabry-P\'erot cavity for modes propagating along the $x$ direction. The solid occupies a thin slab  at $x=0$, and is described by  a minimal model for a two-band semiconductor. The symmetry of the corresponding Wannier orbitals allows for a nonzero dipolar transition matrix element. Along the $y$- and $z$-axis the system is translation invariant, and for the description of the electromagnetic field we include only the modes with polarization along $z$ and wave vector along $x$, because modes with a wave vector $\bm q_{\perp}=0$ and $\bm q_{||}\neq0$ parallel to the  plane of the material are not mixed with the $\bm q_{\perp}$ mode in the semiclassical treatment below. The restriction to one polarization direction can be justified by the symmetry of the orbitals (see below). 

We formulate the Hamiltonian in dipolar gauge, where it can be  decomposed in a number of terms 
\begin{align}
\label{ffaaffaa}
\hat{H} =
\hat{H}_{em}
+
\hat{H}_{0}
+
\hat{H}_{EP}
+
\hat{H}_{PP}
+
\hat{H}_{int},
\end{align}
denoting the empty cavity Hamiltonian $\hat{H}_{em}$, the  free matter Hamiltonian $\hat{H}_{0}$, the screened Coulomb interaction $\hat{H}_{int}$, the light-matter coupling part $\hat{H}_{EP}$, and the dipolar interaction $\hat{H}_{PP}$. These terms will be discussed one-by-one below. The notation is based on Ref.~\onlinecite{Li2020}, which provides a general discussion of tight-binding light-matter Hamiltonians. 

\subsubsection*{Empty cavity Hamiltonian}

The first term in Eq.~\eqref{ffaaffaa}, $\hat{H}_{em}$ describes the electromagnetic field of the cavity.  Throughout this paper we will use natural units (with Lorentz-Heaviside units for electromagnetism) such that $\hbar=1$, $c=1$ and $\epsilon_0=1$.  The free electromagnetic field is thus described by the standard form $\hat{H}_{em}=\frac{1}{2} \int d^3\bm r\Big[ \epsilon(\bm r)^{-1}\hat{\bm{ \Pi}}^2+(\bm \nabla \times  \hat{\bm{A}})^2 \Big]$, where $\hat{\bm{ A}}$ is the transverse vector potential and $\hat{ \bm{\Pi}}$ its canonically conjugate variable, which is given by the transverse component of the displacement field 
\begin{align}
\label{doisplsgfe}
\hat{\bm{\Pi}}(\bm r)=-\hat{ \bm{D}}^T(\bm r).
\end{align}
The space-dependent dielectric function $\epsilon(\bm r)$ accounts for lossless media (such as partially transparent mirrors) which define an arbitrary cavity environment, but does not yet include the 
electromagnetic response of the EI,  which is treated explicitly below. Taking only the modes with $z$-polarization travelling in $x$-direction, the electromagnetic fields are expanded as $\hat{\bm{ A}}(\bm r)=\bm e_z \hat{A}_z(x)/\sqrt{L^2}$, where $L^2$ is the transverse volume. The Hamiltonian then takes the form
\begin{align}
\hat{H}_{em}
&=
\frac{1}{2}
\int d x
\Big[
\frac{1}{\epsilon(x)} 
\hat{\Pi}_z^2(x)+
\big(\partial_x \hat{A}_z(x)\big)^2
\Big],
\label{lwcbas}
\end{align}
where $\hat{A}_z$ and $\hat{\Pi}_z$ are canonically conjugate, $[\hat{A}_z(x),\hat{\Pi}_z(x')]=i\delta(x-x')$.

Below it will be convenient to refer to a discrete mode representation, which, as long as all modes are taken into account, is still general. For $x$-propagating modes with $z$-polarization, the fields are expanded as
\begin{align}
\label{hhhshwh01}
&
\hat{A}_z(x)
=
\sum_{\nu}
\phi_\nu(x) \hat{Q}_{\nu}, 
\\
\label{hhhshwh02}
&
\hat{\Pi}_z(x)
=
\sum_{\nu}
\phi_{\nu}(x)^*
 \epsilon(x)
\,\hat{\Pi}_\nu,
\end{align}
where the operators $\hat{Q}_{\nu}$ and $\hat{\Pi}_\nu$ denote the canonical variables, i.e.,  $[\hat{Q}_\nu,\hat{\Pi}_{\nu'}]=i\delta_{\nu,\nu'}$, and the mode functions satisfy the orthogonality condition $\int dx\,  \epsilon(x)\, {\phi}_{\nu}(x)^* {\phi}_{\nu'}(x) = \delta_{\nu,\nu'}$. These modes may or may not be eigenmodes of the Hamiltonian \eqref{lwcbas}.

\subsubsection*{Uncoupled matter Hamiltonian}
\label{sec:H0}

The EI is described by a two-band tight-binding model with repulsive on-site interaction and a direct band gap. Introducing the spinors
\begin{equation}
\hat{\mathbf{\Psi}}_\mathbf{k}=\begin{pmatrix}
\hat{c}_{\mathbf{k}1}\\
\hat{c}_{\mathbf{k}2}
\end{pmatrix}
\end{equation}
and
\begin{equation}
\hat{\mathbf{\Psi}}_j=\begin{pmatrix}
\hat{c}_{j1}\\
\hat{c}_{j2}
\end{pmatrix}=\frac{1}{\sqrt{N}}\sum_{\mathbf{k}}e^{i\mathbf{k}\cdot\mathbf{R}_j}\hat{\mathbf{\Psi}}_\mathbf{k}
\label{equ:Psi}
\end{equation}
with fermionic annihilation operators $\hat{c}_{\mathbf{k},a}$ ($\hat{c}_{j,a}$) for an electron in band $a$ with quasi-momentum $\mathbf{k}$ (at site $j$) the noninteracting Hamiltonian of the electronic system without light-matter coupling reads
\begin{equation}
\hat{H}_0=\sum_{\mathbf{k}}\epsilon_\mathbf{k}\hat{\mathbf{\Psi}}^\dagger_\mathbf{k}\sigma_z\hat{\mathbf{\Psi}}_\mathbf{k}-\mu\hat{N}.
\label{equ:H}
\end{equation}
Here $\sigma_z$ denotes the $z$-component of the vector of Pauli-matrices $\boldsymbol{\sigma}=(\sigma_x,\sigma_y,\sigma_z)^T$, and the electron-dispersion is taken to be symmetric, with  $\pm\epsilon_\mathbf{k}$ for the lower and upper band, respectively.  In the simulations below, only the density of states $D$ for the energies $\epsilon_\mathbf{k}$ enters. 
We parametrize $\epsilon_\mathbf{k}$ by a continuous variable $\vartheta$, such that
\begin{equation}
\label{stragethetaparemetrization}
\epsilon=\frac{W}{2}\cos\vartheta+\varepsilon_0, 
\end{equation}
where $\vartheta$ takes values between zero and $\pi$, and choose
\begin{equation}
D(\vartheta)=\frac{2}{\pi}\sin^2\vartheta,
\label{equ:dos}
\end{equation}
corresponding to a semi-elliptic density of states  for the variable $\epsilon$ in the range $(\epsilon_0-W/2,\epsilon_0+W/2)$. 
One can expect the results to be qualitatively similar for different $D$, as long as $D$ is regular. 

In addition, $\hat{H}_{int}$ in Eq.~\eqref{ffaaffaa} represents a local electron-electron interaction term 
\begin{equation}
\hat{H}_{int}=U\sum_{j}\hat{n}_{j1}\hat{n}_{j2},
\end{equation}
where $\hat{n}_{j,a}=\hat{c}_{j,a}^\dagger\hat{c}_{j,a}$ is the particle-number operator. The Hamiltonian $H_0+H_{int}$ is symmetric under the $U(1)$ gauge transformation generated by $U(\theta)=e^{i\theta (\hat{N}_1-\hat{N}_2)}$, where $\hat{N}_a=\sum_{j} \hat{c}_{j,a}^\dagger \hat{c}_{j,a}$ is the particle number in band $a$, and therefore the matter Hamiltonian alone conserves the particle number in each band.

\subsubsection*{Dipolar light-matter coupling }

The Hamiltonians $\hat{H}_{EP}$ and $\hat{H}_{PP}$ in Eq.~\eqref{ffaaffaa} denote the light-matter coupling and a dipolar self-interaction, respectively.  In general, this Hamiltonian is derived from the continuum, $\hat{H}_{EP}= \tfrac12\int d^3\bm r\big[ \hat{\bm{\Pi}}(\bm r) \hat{\bm{ P}}(\bm r)+h.c.\big]/\epsilon(\bm r)$ and $\hat{H}_{PP}=\tfrac12\int d^3\bm r\,\hat{\bm{ P}}(\bm r)^2/\epsilon(\bm r)$, where $\hat{\bm{P}}(\bm r)$ is the transverse polarization density.\cite{Li2020} 
In the discrete mode representation \eqref{hhhshwh02}, they read
\begin{align}
\hat{H}_{EP}
&=\frac{1}{2}\sum_{\nu}  \hat{\Pi}_\nu^\dagger \, \hat{P}_\nu + h.c.,
\label{equ:H_EP}\\
\hat{H}_{PP}
&= \frac{1}{2}\sum_{\nu}  \hat{P}_\nu^\dagger  \hat{P}_\nu, 
\label{equ:H_PP}
\end{align}
where the operators $\hat{P}_\nu$ represent the expansion of the 
polarization density
in the same modes as the electric field, Eq.~\eqref{hhhshwh02}. For convenience, we set $\epsilon(x)=1$ at the position $x=0$ of the solid.

Both $\hat{\bm{ P}}$ and $\hat{\bm{ \Pi}}$ must be expanded  in the same set of mode functions if the  expansion is truncated.  In the present case, with a slab of the material at $x=0$ the mode functions are homogeneous over the solid, so that $\hat{P}_{\nu}$ in Eqs.~\eqref{equ:H_EP} and \eqref{equ:H_PP} are  given by $\hat{P}_{\nu}=\phi_{\nu}(0)/\sqrt{L^2} \,\hat{ \mathcal{P}}_z$, where $\hat{ \mathcal{P}}_z$ is the total polarization along $z$. In dipolar gauge, the latter is given by  $\hat { \mathcal{P}}_z= \sum_{ijab} D_{ij}^{ab}c_{i,a}^\dagger c_{j,b} $, with the dipolar transitions matrix elements
\begin{align}
\label{cghjhbjknjkjn}
D_{ij}^{ab}
&=
\int d^3{\bm r}\,
w_{ia}(\bm r)^*
\, z\,
w_{jb}(\bm r)
\end{align} 
between Wannier orbitals $w_{j,a}$. For simplicity we assume that the Wannier orbitals corresponding to the two bands have a symmetry such that a dipolar transition couples only to fields polarized along the $z$-direction (see Fig.~\ref{fig:setup}), and we take into account only the dominant local dipolar matrix elements $q D_{jj}^{12}=q D_{jj}^{21}\equiv g$. Hence, we have $ \hat{\mathcal{P}}_z = g \hat{P}$, where
 \begin{equation}
\hat{P}=\sum_{j}\hat{\mathbf{\Psi}}_j^\dagger\sigma_x\hat{\mathbf{\Psi}}_j=\sum_{\mathbf{k}}\hat{\mathbf{\Psi}}_\mathbf{k}^\dagger\sigma_x\hat{\mathbf{\Psi}}_\mathbf{k}.
\end{equation}

In principle, the light-matter coupling in dipolar gauge also enters all nonlocal ($\mathbf{k}$-dependent) matrix elements in the Hamiltonian via a Peierls substitution.\cite{Li2020} However, the latter can be omitted if the polarization of the field is perpendicular to the direction of the solid (see Fig.~\ref{fig:setup}).

\subsection{Semiclassical description}
\label{sec:sc}
As discussed above, the operators $ \hat{P}_\nu$ and $ \hat{\Pi}_\nu$ are averaged over the solid. In the limit of large $L$, one can therefore use a mean-field decoupling of the light-matter part in Eqs.~\eqref{equ:H_EP} and \eqref{equ:H_PP},
\begin{align*}
	\hat{\Pi}_\nu^\dagger \hat{P}_\nu&\rightarrow \langle \hat{\Pi}_\nu^\dagger \rangle \hat{P}_\nu + \hat{\Pi}_\nu^\dagger \langle \hat{P}_\nu \rangle\\
	\hat{P}_\nu^\dagger  \hat{P}_\nu&\rightarrow \langle \hat{P}_\nu^\dagger \rangle  \hat{P}_\nu + \hat{P}_\nu^\dagger \langle \hat{P}_\nu\rangle.
\end{align*}

Further, for a thin slab of the material with an extent $a$ in $x$-direction that is below a suitable short wavelength cutoff for the light, one can use a mode expansion in which precisely one mode 
($\nu=0$) is non-vanishing over the extent of the solid, with $\phi_0(x)=1/\sqrt{a}$, so that $\int_{solid} dx |\phi_0(x)|^2=1$. Then 
$\hat P_0=g\hat P/\sqrt{V}$.  The decoupling $\hat P_0 \langle\hat P_0 +\hat \Pi_0\rangle$ and $\hat \Pi_0\langle\hat P_0\rangle$ of the relevant interaction terms   leads to the following  mean-field Hamiltonians for light and matter:
\begin{align}
\hat{H}_{mf}^{matter}
&=
\hat{H}_0
+ 
\hat{H}_{int}
-
g 
E_0(t)
\hat{P}
\label{equ:H_sc_matter}
\\
\hat{H}_{mf}^{em}
&=
\hat{H}_{em}
+
\hat \Pi_0\langle\hat P_0\rangle
\end{align}
where $E_0(t)=- \langle\hat P_0 +\hat \Pi_0\rangle/\sqrt{V}$.
The corresponding equations of motion can be recast into the form
\begin{align}
\label{E00000xok}
&E_0(t)=-\frac{\partial_t\langle  \hat{ Q}_0\rangle }{\sqrt{V}}=\frac{-\partial_t A_z(0,t)}{\sqrt{L^2}},
\\
\label{waveequastion}
&\big[\partial_t^2 -\epsilon(x)^{-1}\partial_x^2\big]A_z(x,t)=\sqrt{L^2}  g\dot p(t)\delta(x),
\end{align}
where  $p(t)=\langle \hat{P}\rangle/V$ is the volume averaged polarization, $g\dot p(t)$ 
is the current,  we used $ A_z(0,t)=\hat Q_0/\sqrt{a}$, and $a$ is taken to $0$ in the equation of motion for the fields.

Because the coupling is linear, the equation for the field implies a linear relation between $E_0$ and the polarization $p(t)$ 
  \begin{align}
  \label{ghjs001}
  	E_0(t)&=\int_{0}^{\infty}dt_{r}\,G(t_r)  \,g\,\dot p(t-t_r),
 \end{align}
where the Green's function $G(t)$ depends on the geometry of the cavity. In Fourier space,
 \begin{align}
 	\tilde E_0(\omega)&=-i\omega g\tilde p(\omega)\tilde{G}(\omega).
	\label{fgehe9992}
 \end{align}
If $G(t)$ is known, the problem is reduced to obtaining the time-dependent polarization $p(t)$ from the dynamics of the material, which is determined by the Hamiltonian \eqref{equ:H_sc_matter} with the time-dependent self-consistent field \eqref{ghjs001}. This is equivalent to solving Maxwell's equations coupled to the microscopic dynamics in the material. Below we determine $G(t)$ for the Fabry-P\'erot cavity.

\subsection{Green's function for the Fabry-P\'erot cavity}

To find a solution of Eq.~\eqref{waveequastion} we first consider the free wave equation 
\begin{align}
\label{waveequastion01}
\big(\partial_t^2 -\partial_x^2\big)A_z(x,t)=j_z(t)\delta(x),
\end{align}
which is solved by
\begin{equation}
	A_z(x,t)=\frac{1}{2}\int
	_{-\infty}^{t-|x|}dt'j(t').
\end{equation}
The electric field 
\begin{equation}
	E_z(x,t)=-\partial_tA_z=-\frac{1}{2}j(t-|x|),
	\label{equ:E_z_free}
\end{equation}
allows to write the response of the electric field $E_z(x,t)$ to the current at $x'=0$ in terms of a Green's function $G^{ret}_E(x,t;0,t')=-\frac{1}{2}\delta((t-t')-|x|)$, 
\begin{equation}
\label{fghgffgkkk}
	E_z(x,t)=\int
	_{-\infty}^t dt'\,G^{ret}_E(x,t;0,t')j(t').
\end{equation}
Next, the  effect of lossless dielectric mirrors could be included by introducing a position dependent dielectric constant $\epsilon(x)$ in the wave equation, as in Eq.~\eqref{waveequastion}. Here we regard the mirrors as two infinitely thin layers of material with a finite reflection coefficient $r$ and a transmission coefficient $\tau$ with $\tau^2+r^2=1$. According to Fresnel's equations for perpendicular incidence, the electric field picks up a phase $\pi$  when it is reflected at an interface to an optically thicker medium; therefore, we take the reflection coefficient $r$ to be negative in order to mimic the effect of a dielectric mirror with a higher refractive index than the surrounding medium (Fig.~\ref{fig:setup}). 

Let us imagine an infinitely short current pulse at time zero, $j(t)=\delta(t)$. Without cavity this would give rise to two pulses which 
are released symmetrically from $x=0$ and travel in opposite directions as can be seen from Eq.~(\ref{equ:E_z_free}). If the cavity is included, the electric field is determined by the same equation until the pulses reach the mirrors of the cavity at $t=\frac{L}{2}$. At this point each pulse is partially reflected and transmitted through the mirror, and the reflected and the transmitted pulse must be multiplied by a factor $r$ or $\tau$, respectively. The reflected parts propagate freely through the cavity until they arrive at the second mirror at time $t=\frac{3L}{2}$, where they are again partially reflected and transmitted. The same process may reoccur an infinite number of times so that the electric field at an arbitrary position $x$ inside the cavity (i.e., for $|x|<\frac{L}{2}$) is given by
\begin{equation*}
\begin{split}
E(x,t)=-\frac{1}{2}\Big[
\delta(t-|x|)
&+\sum_{n=1}^{\infty}
r^{n}\delta\big(t-\left(nL-|x|\right)\big)\\
&+
\sum_{n=1}^{\infty}
r^{n}\delta\big(t-\left(nL+|x|\right)\big)\Big].
\end{split}
\end{equation*}
The first delta function in the above expression corresponds to the pulse that has not been reflected yet, whereas the $n$th term in each sum describes a pulse that has been reflected $n$ times. 
The two sums correspond to the two different initial propagation directions. 
With Eq.~\eqref{fghgffgkkk}, this leads to the Green's function
\begin{equation}
	\begin{split}
	G^{ret}_E(x,t;0,t')= &-\frac{1}{2}\delta((t-t')-|x|)\\
	&-\frac{1}{2}\sum_{n=1}^{\infty}\delta((t-t')-(nL-|x|))r^n\\
	&-\frac{1}{2}\sum_{n=1}^{\infty}\delta((t-t')-(nL+|x|))r^n
	\end{split}
\end{equation}
for $|x|<\frac{L}{2}$. The Green's function for the field outside the cavity (i.e., for $|x|>\frac{L}{2}$) can be obtained in a similar way,
\begin{equation}
	G^{ret}_E(x,t;0,t')= -\frac{1}{2} \tau\displaystyle\sum_{n=0}^{\infty}\delta((t-t')-(nL+|x|))r^n.
\end{equation}
The electric field at $x=0$ is given by
 \begin{align}
 	E(0,t)&=\int\limits_{0}^{\infty}dt_{r}G(t_r)j(t-t_r),
 	\label{equ:E0_integr}
	\\
  G(t)&=-\frac{1}{2}\delta(t)-\sum_{n=1}^{\infty}\delta(t-nL)r^n,	
 \end{align}
 where   $G(t-t')=G^{ret}_E(0,t;0,t')$. 
 In Fourier space,
 \begin{align}
 	\tilde{G}(\omega)&=	-\frac{1}{2}\left(\frac{1+e^{i\omega L}r}{1-e^{i\omega L}r}\right).
 \end{align}
 Applying this solution to \eqref{E00000xok} and \eqref{waveequastion}, we finally get Eqs.~\eqref{ghjs001} and \eqref{fgehe9992}.

\subsection{Mean-field decoupling of the electron-electron interaction}
\label{sec:mf-decoupl}
We will treat the local electron-electron interaction   $\hat{H}_{int}$ in the Hamiltonian \eqref{equ:H_sc_matter} using a mean-field decoupling of the form
\begin{equation}
\hat{n}_{j1}\hat{n}_{j2}\rightarrow n_{j1}\hat{n}_{j2}+\hat{n}_{j1}n_{j2}-\phi^*\hat{c}_{j2}^\dagger \hat{c}_{j1}-\phi\hat{c}_{j1}^\dagger\hat{c}_{j2},
\end{equation}
where $n_a=\langle\hat{n}_{j,a}\rangle$ represents the average number of electrons in band $a$ per lattice site, and 
$\phi$ is the order parameter of the excitonic condensate (which is the same for all lattice sites $j$)
\begin{align}
	\phi=&\langle \hat{c}_{j2}^\dagger \hat{c}_{j1} \rangle=\langle \hat{c}_{j1}^\dagger \hat{c}_{j2} \rangle^*.
\end{align} 
The mean-field Hamiltonian is thus given by
\begin{equation}
\hat{H}_{mf}=\sum_{\mathbf{k}}\hat{\mathbf{\Psi}}_\mathbf{k}^\dagger h_\mathbf{k}\hat{\mathbf{\Psi}}_\mathbf{k}\\
\label{equ:H_mf}
\end{equation}
with the single particle Hamiltonian 
\begin{equation}
h_\mathbf{k}=\begin{pmatrix}
\epsilon_\mathbf{k}+Un_2-\mu&-U\phi-gE_0(t)\\
-U\phi^*-gE_0(t)&-\epsilon_\mathbf{k}+Un_1-\mu
\end{pmatrix}.
\label{equ:h_k1}
\end{equation}
Henceforth, we set the chemical potential to $\mu=U/2$ and fix the average electron number per lattice site to $n_{tot}=n_1+n_2=1$. With this, (\ref{equ:h_k1}) becomes 
\begin{equation}
h_\mathbf{k}=\begin{pmatrix}
\epsilon_\mathbf{k}+\frac{1}{2}U(n_2-n_1)&-U\phi-gE_0(t)\\
-U\phi^*-gE_0(t)&-\epsilon_\mathbf{k}-\frac{1}{2}U(n_2-n_1)
\end{pmatrix}.
\label{equ:h_k2}
\end{equation}
Within mean-field approximation the state of the material is described by the momentum-dependent density matrix
\begin{equation}
	\rho_\mathbf{k}=\begin{pmatrix}
		\langle\hat{c}^\dagger_{\mathbf{k}1}\hat{c}_{\mathbf{k}1}\rangle & \langle\hat{c}^\dagger_{\mathbf{k}2}\hat{c}_{\mathbf{k}1}\rangle\\
		\langle\hat{c}^\dagger_{\mathbf{k}1}\hat{c}_{\mathbf{k}2}\rangle & \langle\hat{c}^\dagger_{\mathbf{k}2}\hat{c}_{\mathbf{k}2}\rangle
	\end{pmatrix}
	\label{equ:rho_k}
\end{equation}
and the corresponding local density matrix
\begin{equation}
	\rho=\begin{pmatrix}
		n_1 & \phi\\
		\phi^* & n_2
	\end{pmatrix}
	=
	\frac{1}{N}\sum_{\mathbf{k}}\rho_\mathbf{k}.
	\label{equ:rho_from_rho_k}
\end{equation}
The expectation value $\rho_\mathbf{k}$ is obtained from the self-consistent mean-field Hamiltonian $\hat{H}_{mf}$. Diagonalizing $h_\mathbf{k}=W_\mathbf{k}D_\mathbf{k}W_\mathbf{k}^\dagger$ with a diagonal matrix $D_\mathbf{k}=\text{diag}(E_\mathbf{k}^{(+)},E_\mathbf{k}^{(-)})$, leads to the expression 
\begin{equation}
\rho_\mathbf{k}=W_\mathbf{k}\begin{pmatrix}
f(E_\mathbf{k}^{(+)})&0\\
0&f(E_\mathbf{k}^{(-)})
\end{pmatrix}W_\mathbf{k}^\dagger,
\label{equ:res_rho_k}
\end{equation}
with the Fermi-function $f(x)=\frac{1}{1+e^{\beta x}}$. With this, the static solution for the local density matrix $\rho$ can be obtained self-consistently from Eq.~(\ref{equ:res_rho_k}) and (\ref{equ:rho_from_rho_k}). 

\subsection{Pseudospin representation}
\label{sec:pseudospin}

To study the collective modes of the EI, it is convenient to adopt an Anderson pseudospin representation. We define a three-component pseudospin-vector 
\begin{equation}
\hat{\mathbf{s}}_\mathbf{k}=\begin{pmatrix}
\hat{s}_\mathbf{k}^x\\
\hat{s}_\mathbf{k}^y\\
\hat{s}_\mathbf{k}^z
\end{pmatrix},\text{  }\hat{s}_\mathbf{k}^\alpha=\frac{1}{2}\hat{\mathbf{\Psi}}_\mathbf{k}^\dagger\sigma_\alpha\hat{\mathbf{\Psi}}_\mathbf{k},
\end{equation}
which satisfies the spin algebra. With the three-component vector
\begin{equation}
\mathbf{\Phi}=\frac{1}{N}\sum_\mathbf{k}\mathbf{s}_\mathbf{k}=\begin{pmatrix}
	\phi'\\
	\phi''\\
	m
	\end{pmatrix}
\end{equation}
[$\phi'=\operatorname{Re}\{\phi\}$, $\phi''=-\operatorname{Im}\{\phi\}$, $m=\frac12(n_1-n_2)$], the mean-field Hamiltonian reads
\begin{equation}
\hat{H}_{mf}=\sum_\mathbf{k}\mathbf{B}_\mathbf{k}\cdot\hat{\mathbf{s}}_\mathbf{k},
\label{equ:H_mf_spin}
\end{equation}
where
\begin{equation}
\mathbf{B}_\mathbf{k}=2
\begin{pmatrix}
-gE_0(t)\\
0\\
0
\end{pmatrix}
+2
\begin{pmatrix}
0\\
0\\
\epsilon_\mathbf{k}
\end{pmatrix}
-2U\bm \Phi
\label{equ:B_k}
\end{equation}
is the pseudomagnetic field. The Heisenberg equation of motion  is given by the pseudospin precession,
\begin{equation}
\frac{d\hat{s}_\mathbf{k}^\alpha}{dt}
=i\left[\hat{H}_{mf},\hat{s}_\mathbf{k}^\alpha\right]
=\mathbf{B}_\mathbf{k}\times\hat{\mathbf{s}}_\mathbf{k}.
\label{equ:scEOM_spin}
\end{equation}

Finally, the expression of the polarization operator in terms of the pseudospin variables is given by $\hat{P}=2\sum_{\mathbf{k}}\hat{s}_\mathbf{k}^x$, so that 
$\langle \hat P\rangle=2N\phi'$, and therefore $p(t)=\frac{N}{V}2\phi'(t)$. Henceforth we will set the factor $\frac{N}{V}$ to one  (fixing the volume of the unit cell), so that
\begin{equation}
p(t)=2\phi'(t).
\end{equation}

\subsection{Linear susceptibility}
\label{sec:suscept}

To calculate the collective modes, we derive the linear susceptibility of $\phi'$, $\phi''$ and $m$ to an external field $\mathbf{f}\propto e^{-i\omega t}$ that couples to the pseudospin operators like
\begin{equation}
\hat{H}=\hat{H}_{mf}+\mathbf{f}\cdot\sum_{\mathbf{k}}\hat{\mathbf{s}}_\mathbf{k},
\end{equation}
where $\hat{H}_{mf}$ is still given by Eq.~(\ref{equ:H_mf_spin}). 
We define the full linear susceptibility $\boldsymbol{\chi}$ such that
\begin{equation}
	\delta\mathbf{\Phi}=\boldsymbol{\chi}\mathbf{f}.
	\label{equ:chi_def}
\end{equation}
The external field $\mathbf{f}$ causes a change in $\mathbf{B}_\mathbf{k}$ as it alters the values of $\phi',\phi'',m$ and $E_0$. We first define the  bare response
\begin{equation}
\delta\mathbf{s}_\mathbf{k}=\bm \chi_\mathbf{k}^0\mathbf{f}
\label{equ:bare_sus_def}
\end{equation}
of an individual pseudospin $\langle\hat{\mathbf{s}}_\mathbf{k}\rangle$ at fixed  $\mathbf{B}_\mathbf{k}\equiv\mathbf{B}_\mathbf{k}^{(0)}$, where $\mathbf{B}_\mathbf{k}^{(0)}$ still denotes the value of $\mathbf{B}_\mathbf{k}$ evaluated at $\mathbf{f}=0$. In analogy to Eq.~(\ref{equ:scEOM_spin}) the new semi-classical equation of motion reads
\begin{equation}
\dot{\mathbf{s}}_\mathbf{k}=(\mathbf{B}_\mathbf{k}^{(0)}+\mathbf{f})\times\mathbf{s}_\mathbf{k},
\end{equation}
with the shorthand notation $\mathbf{s}_\mathbf{k}\equiv\langle \hat{\mathbf{s}}_\mathbf{k}\rangle$ and $s_\mathbf{k}^\alpha\equiv\langle\hat{s}_\mathbf{k}^\alpha\rangle$.
Rewriting the expectation value of the pseudospin as
\begin{equation}
\mathbf{s}_\mathbf{k}=\mathbf{s}_\mathbf{k}^{(0)}+\delta\mathbf{s}_\mathbf{k},
\end{equation}
where $\mathbf{s}_\mathbf{k}^{(0)}$ denotes the static expectation value evaluated at $\mathbf{f}=0$,
and taking $\delta\mathbf{s}_\mathbf{k}\propto e^{-i\omega t}$, one finds
\begin{equation}
-i\omega\delta\mathbf{s}_\mathbf{k}=\mathbf{f}\times\mathbf{s}_\mathbf{k}^{(0)}+\mathbf{B}_\mathbf{k}^{(0)}\times\delta\mathbf{s}_\mathbf{k}.
\end{equation}
This can be rewritten as
\begin{equation}
(i\omega+\mathbf{B}_\mathbf{k}^{(0)}\times)\delta\mathbf{s}_\mathbf{k}=\mathbf{s}_\mathbf{k}^{(0)}\times\mathbf{f},
\end{equation}
or
\begin{equation}
A_\mathbf{k}\delta\mathbf{s}_\mathbf{k}=C_\mathbf{k}\mathbf{f},
\label{equ:ds_mat}
\end{equation}
with the $3\times3$ matrices 
\begin{equation}
A_\mathbf{k}\equiv i\omega\mathbbm{1}_{3\times 3}+\begin{pmatrix}
0& -B_{\mathbf{k}z}^{(0)}& B_{\mathbf{k}y}^{(0)}\\
B_{\mathbf{k}z}^{(0)}&0& -B_{\mathbf{k}x}^{(0)}\\
-B_{\mathbf{k}y}^{(0)}& B_{\mathbf{k}x}^{(0)}& 0
\end{pmatrix}
\end{equation}
and
\begin{equation}
C_\mathbf{k}=\begin{pmatrix}
0 & -s_{\mathbf{k}z}^{(0)} & s_{\mathbf{k}y}^{(0)}\\
s_{\mathbf{k}z}^{(0)} & 0 & -s_{\mathbf{k}x}^{(0)}\\
-s_{\mathbf{k}y}^{(0)}& s_{\mathbf{k}x}^{(0)} &0
\end{pmatrix}.
\end{equation}
Comparing Eqs.~(\ref{equ:bare_sus_def}) and (\ref{equ:ds_mat}) one finds that 
\begin{equation}
\chi_\mathbf{k}^0=A_\mathbf{k}^{-1}C_\mathbf{k}.
\end{equation}
 
\begin{figure*}
	\centerline{
	\includegraphics[width=\textwidth]{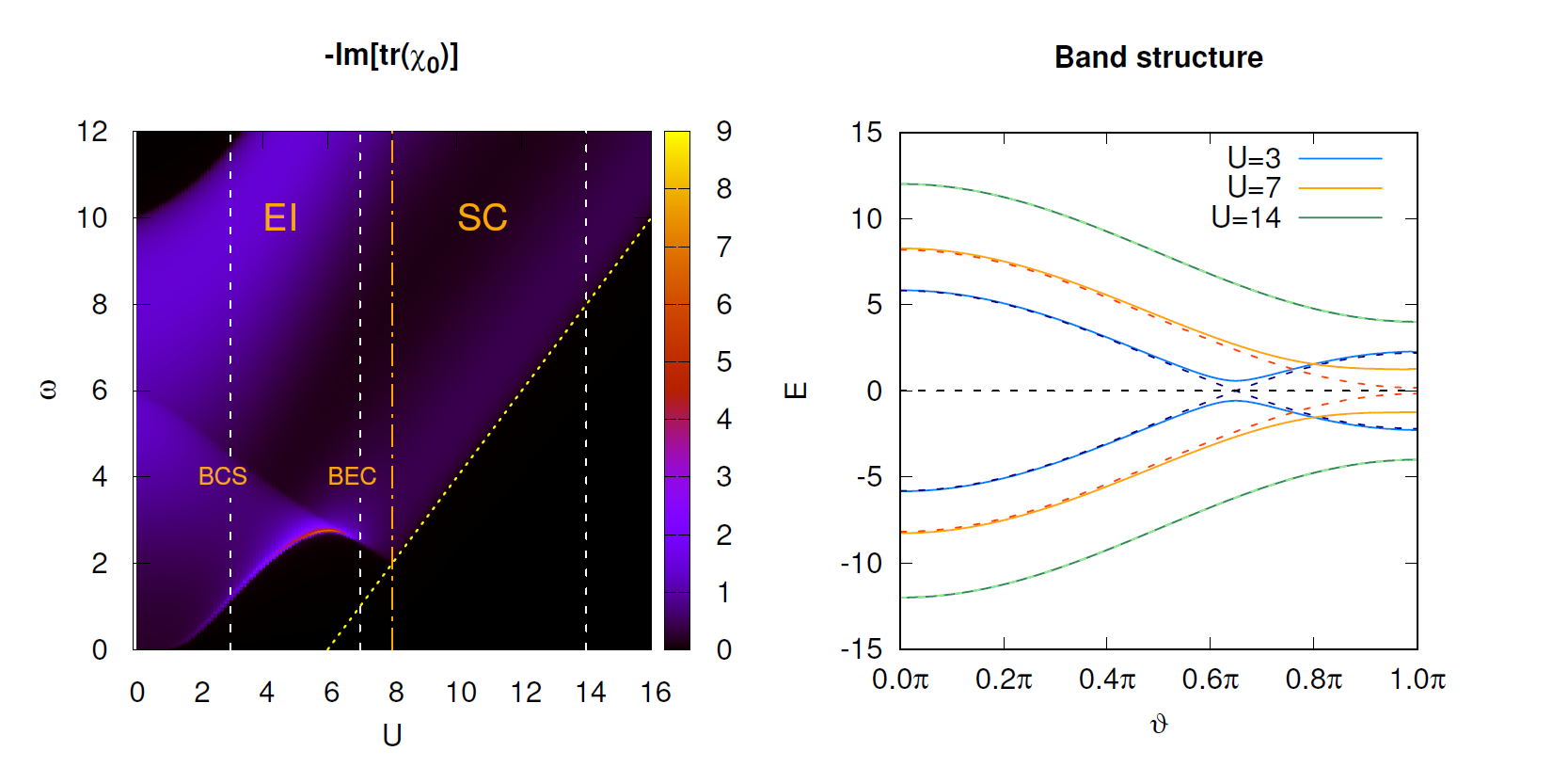}}
	\caption{ Left panel: Imaginary part of $-\operatorname{tr}\{\boldsymbol{\chi}_0\}$ as a function of the frequency $\omega$ and the electron-electron interaction $U$. White dashed lines mark the interactions for which the energy bands are shown in the right panel. Labels indicate the BCS and the BEC regime of the EI phase, as well as the SC (normal state) for $U>8$. Right panel: Electron dispersion of the interacting system at $U=3$, $U=7$ and $U=14$. The dashed lines indicate the non-interacting bands plus Hartree-shift.  Energies are parametrized by the variable $\vartheta$, see Eq.~\eqref{stragethetaparemetrization}.	}
	\label{fig:bands_bareSus}
\end{figure*}
In the next step we include the feedback on $\mathbf{B}_\mathbf{k}$,
\begin{equation}
\mathbf{B}_\mathbf{k}=\mathbf{B}_\mathbf{k}^{(0)}+\delta\mathbf{B}_\mathbf{k}.
\end{equation}
Then the change in $\mathbf{s}_\mathbf{k}$ can be calculated from the  bare response using the expression  
\begin{equation}
\delta\mathbf{s}_\mathbf{k}=\chi_\mathbf{k}^{0}\left[\mathbf{f}+\delta\mathbf{B}_\mathbf{k}\right].
\label{equ:ds_k}
\end{equation}
From the definition of $\mathbf{B}_\mathbf{k}$ [c.~f.~Eq.~(\ref{equ:B_k})] it follows that
\begin{equation}
\delta\mathbf{B}_\mathbf{k}=-2U
\delta \bm \Phi
-\begin{pmatrix}
1\\
0\\
0
\end{pmatrix}2g\delta E_0.
\label{equ:dBfromdE}
\end{equation}
Since the above expression does not depend on $\mathbf{k}$, we can omit the $\mathbf{k}$-index, i.e., we define $\delta\mathbf{B}\equiv\delta\mathbf{B}_\mathbf{k}$.  With \eqref{fgehe9992}, we have
$\delta E_0=-i\omega\tilde{G}(\omega)2g\delta\phi'$,
and therefore
\begin{equation}
\label{eq:responseB}
	\delta\mathbf{B}=V\delta\mathbf{\Phi}
\end{equation}
with the interaction matrix
\begin{equation}
 V=V_U+V_g,
\end{equation}
where
\begin{equation}
 V_U=-2U\mathbbm{1}_{3\times3}
 \label{equ:VU}
\end{equation}
and
\begin{equation}
V_g=4g^2i\omega\tilde{G}(\omega)\begin{pmatrix}
1 & 0 & 0\\
0 & 0 & 0\\
0 & 0 & 0 
\end{pmatrix}.
\label{equ:Vg}
\end{equation}

In order to derive an expression for 
the full linear susceptibility \eqref{equ:chi_def}
we multiply Eq.~(\ref{equ:ds_k}) by $\frac{1}{N}$ and sum it over all $\mathbf{k}$-values.
This yields the following equation
\begin{equation}
	\delta\mathbf{\Phi}=\boldsymbol{\chi}_{0}\left[\mathbf{f}+\delta\mathbf{B}\right]=\boldsymbol{\chi}_{0}\left[\mathbf{f}+V\delta\mathbf{\Phi}\right],
	\label{equ:dPhi_chi0}
\end{equation}
where we have defined the bare susceptibility
\begin{equation}
	\boldsymbol{\chi}_{0}=\frac{1}{N}\sum_{\mathbf{k}}\chi_\mathbf{k}^{0}.
	\label{equ:bare_sus}
\end{equation}
Solving Eq.~(\ref{equ:dPhi_chi0}) for $\delta\mathbf{\Phi}$ and comparing the result to Eq.~(\ref{equ:chi_def}) finally leads to the RPA equation
\begin{equation}
	\boldsymbol{\chi}=\left[\mathbbm{1}_{3\times 3}-\boldsymbol{\chi}_{0}V\right]^{-1}\boldsymbol{\chi}_{0}.
	\label{equ:full_sus}
\end{equation}

\section{Results}
\label{sec:results}

\subsection{Equilibrium state}
\label{sec:res_equ}

In this section we study the equilibrium states and collective modes of the EI at different values of the electron-electron interaction $U$. We first recapitulate the equilibrium phase diagram of the model. For all  calculations the bandwidth is   $W=8$ and the bare energy shift of the bands is $\varepsilon_0=1$. Therefore,  at $U=0$ the material is a normal metal with overlapping bands. The right panel of Fig.~\ref{fig:bands_bareSus} shows the mean-field band structure  for three different values of $U$. Dashed lines  indicate the energy bands in the normal state. For $U>0$, the Hartree term $\frac{1}{2}U(n_2-n_1)$ in the diagonal entries of the single particle Hamiltonian (\ref{equ:h_k2}) results in a shift. With increasing $U$, this opens a gap, leading to a transition from a metal to a semiconductor (SC) with a fully occupied conduction band and an empty valence band. In the SC regime, the occupation difference is thus fixed to $n_1-n_2=1$ so that the band gap increases linearly with $U$. 

For interaction  $0<U<8$ the excitonic order parameter takes a finite value at zero temperature. This gives rise to  off-diagonal terms $U\phi$ in the mean-field Hamiltonian, which open a band-gap at the crossing points of the Hartree-shifted bands. Even though the EI phases at different $U$ are continuously related, the mechanism for the transition differs when the EI is approached from the SC or the 
metallic state, and the metal-EI-SC transition constitutes a BCS-BEC crossover scenario:\cite{Zenker2012}  Starting from the metallic phase, the formation of the exciton condensate can be described by a BCS-like process, while the transition from the SC to the EI state is due to a softening of the excitonic mode, which can be interpreted as a  Bose-Einstein condensation (BEC) of excitons. 

Importantly, it turns out that the self-consistent equilibrium state is entirely independent of the light-matter coupling. This fact will be analyzed in more detail in Sec.~\ref{sec:el-phon}. In short, from Eq.~\eqref{ghjs001}, one can see that the field $E_0$, which is the only feedback of the electromagnetic field on the matter, vanishes whenever the system is time-independent, and hence $\dot p=0$. As a consequence, the equilibrium state does not depend on the light-matter coupling strength $g$ and can be evaluated at $g=0$. Also, the phase $\theta$ of the order parameter $\phi=|\phi |e^{i\theta}$ in the equilibrium solution is arbitrary. It will be fixed to  $\theta=0$ unless stated otherwise. 

The left panel of Fig.~\ref{fig:bands_bareSus} shows the imaginary part $-\operatorname{tr}\{\boldsymbol{\chi}_0\}$ of the bare response $\chi_0$ of the system, which is given by Eq.~(\ref{equ:bare_sus}).  To
analyze the susceptibility matrix $\boldsymbol{\chi}$ it is convenient to compute its trace, which is equal to the sum of its eigenvalues and therefore  contains all the relevant information on the excitation 
spectrum. The region of non-vanishing susceptibility marks the particle-hole continuum, which is limited by the possible transitions between valence and conduction band. 
Its upper and lower boundary at a given value of $U$ is given by the maximum and minimum band gap, respectively. 
The phase boundary between the EI and the SC state at $U=8$ is indicated by the orange dotted-dashed line. It can be seen clearly how the $U$-dependence of the minimum band-gap changes at the transition. The yellow dotted line shows the minimum band gap for the bare bands (non-interacting electron dispersion plus Hartree-shift).  
It  coincides with the lower edge of the particle-hole continuum if the system is in the SC state.  Another interesting feature is the diagonal kink in the particle-hole continuum, which starts at about $\omega=6$  for $U=0$ and ends at the BEC-BCS crossover, where it reaches the lower boundary of the particle hole-continuum. As can be seen for $U=3$ in the right panel of Fig.~\ref{fig:bands_bareSus},  the separation of the bands on the BCS side of the crossover has a global minimum  at $0<U<\pi$ and a local minimum at $\vartheta = \pi$. The latter gives rise to a kink in the particle-hole continuum at the corresponding frequency $\omega$. 

\subsection{Linear susceptibility and collective modes}
\label{sec:res_linSus}

\begin{figure}
	\centering
	\includegraphics[width=0.45\textwidth]{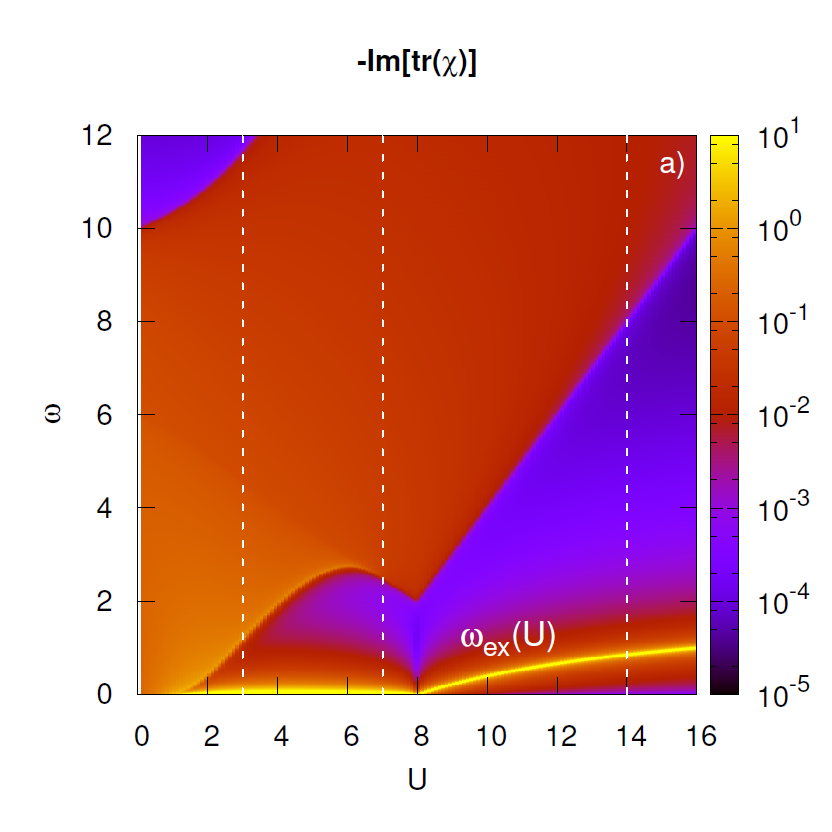} \\
	\includegraphics[width=0.45\textwidth]{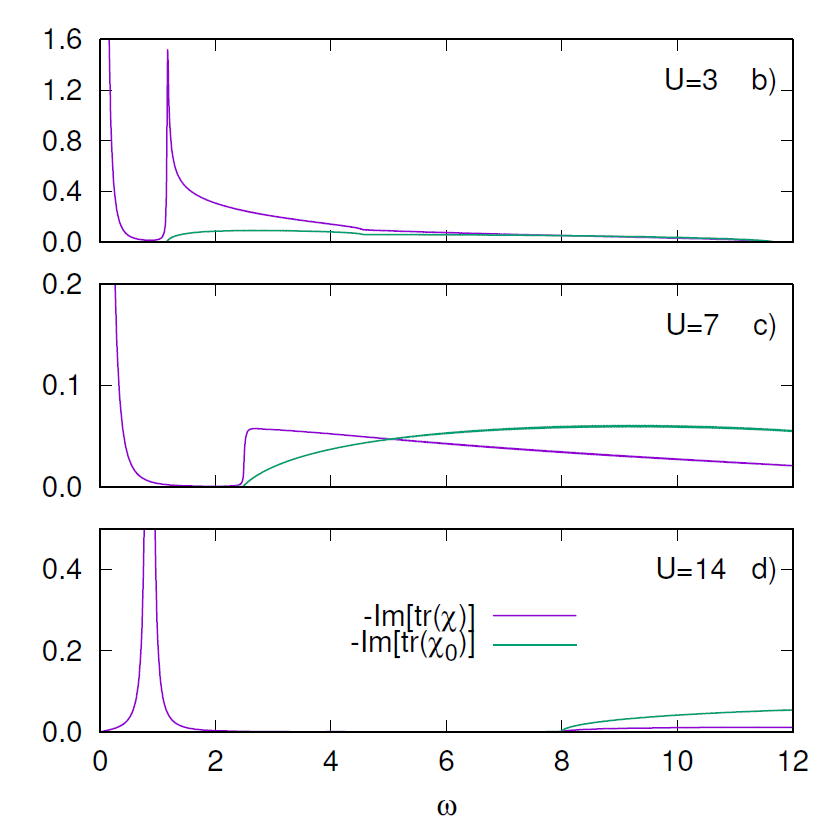}
	\caption{a) Imaginary part of $-\operatorname{tr}\{\boldsymbol{\chi}\}$ for the material without light-matter interaction (i.e., $g=0$).  A small imaginary part has been  added to the frequency in order to make $\delta$-peaks visible. The white dashed lines in panel a) indicate the values of $U$ for the plots in the lower panels.  b)-d) Imaginary part of $-\operatorname{tr}\{\boldsymbol{\chi}\}$ and  $-\operatorname{tr}\{\boldsymbol{\chi}_0\}$ as a function of the frequency for three different values of $U$. }\label{fig:sus_g0}
\end{figure}

In the following we will consider the full susceptibility $\boldsymbol{\chi}$, which, in contrast to $\bm \chi_0$, does depend on the light-matter coupling $g$. 
We will start by recapitulating the response of the bare EI ($g=0$). 
Fig.~\ref{fig:sus_g0} shows the imaginary part of $-\operatorname{tr}\{\boldsymbol{\chi}\}$. The lower panels Fig.~\ref{fig:sus_g0}b-d display three slices through Fig.~\ref{fig:sus_g0}a at different values of $U$, as indicated by the white dashed lines. In addition to the particle-hole continuum, which is already contained in the bare-susceptibility, one can identify the collective excitations. They  correspond to zero eigenvalues of the matrix $[1-\boldsymbol{\chi}_0V]$ [c.f.~Eq.~\eqref{equ:full_sus}], and appear as sharp peaks in $-\operatorname{tr}\{\boldsymbol{\chi}\}$: In the normal state  ($U>8$), there is a pole at some frequency $\omega_{ex}(U)>0$ within the band gap. This mode softens at the transition to the EI phase at $U=8$, which indicates the Bose-Einstein condensation of excitons. For $U<8$ there is a sharp peak at $\omega=0$. 
The nature of these modes is seen from the eigenvectors of $\bm \chi$ at the corresponding resonance frequencies. The eigenvector of the $\omega=0$ pole in the EI phase is proportional to $(0,1,0)^T$.
Because the equilibrium order parameter is chosen to be real, this mode is therefore identified as the phase mode. 
In the normal phase, the eigenvector corresponding to the exciton mode is proportional to $(1,i,0)^T$ for $\omega>0$ and $(1,-i,0)^T$ for $\omega<0$, which implies a circular oscillation of $\phi$ around the equilibrium value $\phi=0$.  Furthermore, the line-outs in Fig.~\ref{fig:sus_g0}b-d show the amplitude mode of the condensate. For small $U$ (on the BCS side of the phase diagram), the amplitude mode corresponds to a zero eigenvalue of $[1-\boldsymbol{\chi}_0V]$ at the lower edge of the particle hole continuum, where the mode appears as a sharp peak ($U=3$, Fig.~\ref{fig:sus_g0}b). At the resonance frequency, the eigenvector of $\boldsymbol{\chi}$ with the largest absolute value is approximately proportional to $(1,0,0)^T$, i.e., it points in radial direction. With this the corresponding excitation can be identified as the  amplitude mode of the excitonic order parameter.  For larger $U$, the mode becomes strongly broadened ($U=7$, Fig.~\ref{fig:sus_g0}c), since the zero eigenvalue of $[1-\boldsymbol{\chi}_0V]$ is shifted into the particle-hole continuum, and it has disappeared at $U=14$ (Fig.~\ref{fig:sus_g0}d).

\begin{figure}
	\centering
	\includegraphics[width=0.45\textwidth]{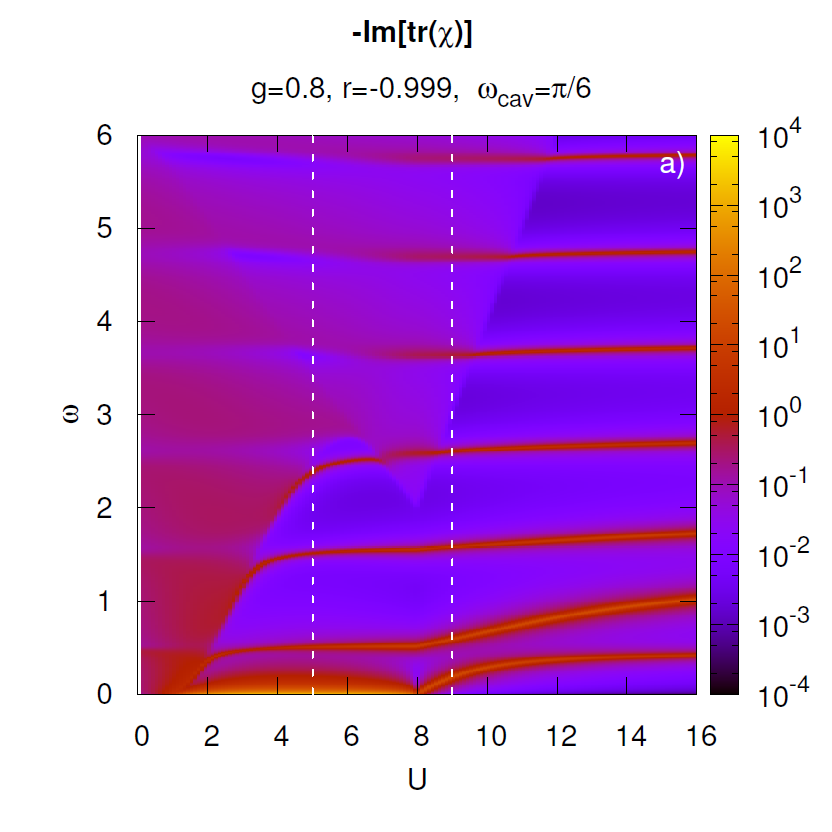}\\
	\includegraphics[width=0.45\textwidth]{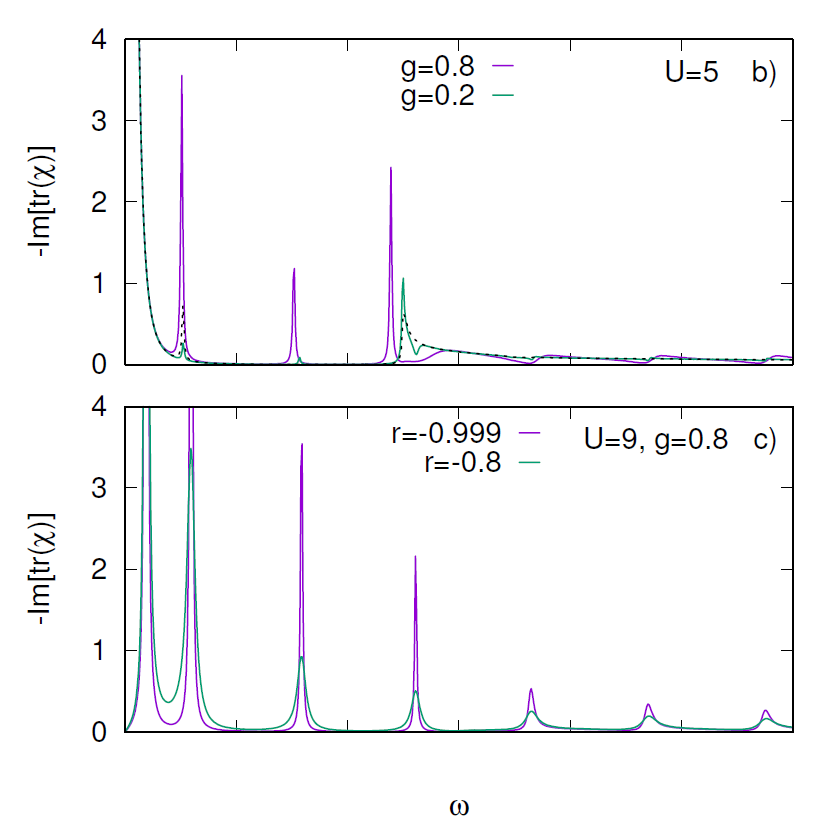}
	\caption{a) Imaginary part of $-\operatorname{tr}\{\boldsymbol{\chi}\}$ for coupling $g=0.8$, reflection coefficient of $r=-0.999$, and cavity frequency $\omega_{cav}=\pi/6$. b) Line-out of the data at $U=5$, for weak and strong coupling $g$. The dashed black line is obtained from a one-mode approximation at $\tilde g=0.2$ (see Section.~\ref{sec:el-phon}). c) Line-out of data at $U=9$, including data for a low reflectivity $r=-0.8$ of the cavity mirrors.}
\label{fig:sus_gnonz}
\end{figure}

We now turn to the case $g>0$, to analyze how the light-matter interaction influences the collective excitations of the system (Fig.~\ref{fig:sus_gnonz}). The boundaries of the particle-hole continuum remain unchanged, because the latter is already contained in the $g$-independent bare susceptibility $\boldsymbol{\chi}_0$. However, there are additional resonances at odd integer multiples of the cavity frequency $\omega_{cav}=\frac{\pi}L$. Resonances at even multiples of $\omega_{cav}$ do not appear, because the corresponding frequency component of the electric field has a node at the center of the cavity, where the material is located. Comparing the purple line for $r=-0.999$ and the green line for $r=-0.8$ in Fig.~\ref{fig:sus_gnonz}c shows that a lower reflectivity of the mirrors leads to a stronger damping of the cavity modes. Moreover, one can observe various effects that occur when the cavity resonances hybridize with other parts of the spectrum: Firstly, a hybridization of the cavity modes and the particle-hole continuum  gives rise to an asymmetric line shape with a dip at lower frequencies and a maximum at higher frequencies, which resembles a Fano-resonance. Secondly, whenever a cavity mode intercepts the exciton peak in the insulating phase there is an avoided crossing, as can be seen around $U=9,\omega=0.5$ in Fig~\ref{fig:sus_gnonz}a. Lastly, the amplitude mode in the EI phase can be pushed out of the particle hole continuum by a cavity mode in its proximity if the light-matter coupling is strong enough (see the behavior around $\omega\approx 2.5$ and $U=5$ in Fig~\ref{fig:sus_gnonz}a). 

The phase mode in the EI phase, however, always remains at $\omega=0$, even though the light-matter interaction breaks the $U(1)$ symmetry of the system. As mentioned in the introduction, this is in stark contrast to the coupling of the $U(1)$ invariant two-band model to a generic oscillator mode, which adds a mass to the phase mode.  To elucidate the origin of this behavior, it is helpful to contrast the model studied above directly with a model of an EI with a generic coupling to a phonon mode (electron-phonon coupling). A one-to-one comparison is  facilitated by restricting the field in the cavity to a single mode, as  discussed in greater detail in the following section.

\subsection{Single-mode models}
\label{sec:el-phon}

\subsubsection{Model}

In model calculations, it is convenient to replace the field in the cavity by one or a few modes. To derive the corresponding Hamiltonian, the modes $\hat{\Pi}_\nu$ and $\hat{Q}_\nu$ can be taken as normal modes of the resonator Hamiltonian $H_{em}$ [Eq.~\eqref{lwcbas}] so that $\hat{H}_{em} = \sum_\nu \frac{1}{2} (\hat{\Pi}_\nu^2 +\omega_\nu^2 \hat{Q}_\nu^2 )$. If we keep only a single mode $\nu=0$, the total Hamiltonian reads
\begin{align}
\label{sffsddsdsds}
\hat{H}
=
\hat{H}_0
+ \hat{H}_{int}
+
\frac{1}{2} \Big[ 
\Big(
\hat{\Pi}_0
+
\frac{\tilde g}{\sqrt{L^2}}
\hat{P}
\Big)^2+\omega_0^2
\hat{Q}_0^2
\Big],
\end{align}
with $\tilde g=\phi_0(0)g$.  We will see that this Hamiltonian has a similar phenomenology as the full cavity Hamiltonian,  in particular a massless phase mode. In an energy range where the behavior of the system is dominated by a single mode, the one-mode approximation 
is quantitatively accurate: For example, the dashed black line in Fig.~\ref{fig:sus_gnonz}b shows the spectrum of the one-mode Hamiltonian for the same cavity frequency as the full Hamiltonian and a coupling strength of $\tilde{g}=0.2$. The spectrum captures well the 
low-energy behavior around the mode frequency, while the broadening due to the cavity loss ($|r|<1$) as well as  the signatures of the higher cavity modes are missing. At larger coupling, the different cavity modes are less well separated, and the one-mode approximation becomes quantitatively less accurate. For the following qualitative discussion of the low frequency behavior, however, the one-mode approximation will be sufficient.

We contrast the one-mode cavity Hamiltonian with the coupling to a generic oscillator, such as an optical phonon. The excitonic insulator with electron-phonon coupling is obtained by replacing $\hat{H}_{EP}$ and $\hat{H}_{PP}$ in Eq.~(\ref{ffaaffaa}) by a  Holstein interaction
\begin{equation}
\hat{H}_{coupl}^{ph}=g_{ph}\sum_{j}(\hat{b}_j^\dagger+\hat{b}_j)(\hat{c}_{j2}^\dagger\hat{c}_{j1}+\hat{c}_{j1}^\dagger\hat{c}_{j2}),
\end{equation}
where the operator $\hat{b}_j^\dagger$ ($\hat{b}_j$) is the creation (annihilation) operator of an Einstein phonon at site $j$, and $g_{ph}$ determines the electron-phonon coupling strength. The free phonon Hamiltonian is given by $\hat{H}_{free}^{ph}=\omega_0\sum_{j}(\hat{b}_j^\dagger\hat{b}_j+\tfrac12)$. We treat the model within mean-field theory, in analogy to the decoupling of the light-matter interaction. In mean-field approximation, only the homogeneous ($q=0$) phonon mode $\hat b_{0} = \frac{1}{\sqrt{L^2}} \sum_j \hat b_j$ is relevant. Singling out the $q=0$ mode, the Hamiltonian becomes
\begin{align}
\label{sffsddsdsds01}
\hat{H}
=
\hat{H}_0
+
 \hat{H}_{int}
+
\frac{g_{ph}\sqrt{2\omega_0}}{\sqrt{L^2}}
\hat{Q}_{ph} \hat P
+
\hat H_{free}^{ph},
\end{align}
where $\hat{Q}_{ph}=(\hat b_{0}^\dagger+\hat b_{0})/\sqrt{2\omega_0}$, $\hat{\Pi}_{ph}$ is the canonically conjugate momentum, $\hat H_{free}^{ph}=\frac{1}{2} \big(\hat{\Pi}_{ph}^2+\hat{Q}_{ph}^2\omega_{0}^2\big)$, and terms involving $q\neq0$ phonons are omitted. With a canonical transformation $\hat\Pi_{0}=\hat{Q}_{ph}\omega_{0}$, $\hat Q_0=-\hat \Pi_{ph}/\omega_{0}$ the electron-phonon interaction is transformed to $\frac{\tilde g}{\sqrt{L^2}}\hat{\Pi}_{0} \hat P$, with $\tilde g=g_{ph}\sqrt{2/\omega_{0}}$. Thus, the Hamiltonian is entirely analogous to the single mode cavity Hamiltonian \eqref{sffsddsdsds}, apart from the absence of a term proportional to $\hat P^2$. We therefore define 
\begin{align}
\label{sffsddsdsds03}
\hat{H}(\alpha) 
=
\hat{H}_{0}
+
 \hat{H}_{int}
+
\frac{\tilde g}{\sqrt{L^2}}
\hat{\Pi}_{0} \hat P
+
\hat H_{free}^{osc} 
+
\alpha
\frac{\tilde g^2  \hat P^2}{2L^2},
\end{align}
where 
$\hat H_{free}^{ocs}=\frac{1}{2} \big(\hat{\Pi}_{0}^2+\hat{Q}_{0}^2\omega_{0}^2\big)$.
Furthermore, we have introduced the factor $\alpha$ such that $\alpha=0$ corresponds to the phonon Hamiltonian, and $ \alpha=1$ yields the one-mode cavity Hamiltonian. 

\begin{figure*}
	\centering
	\includegraphics[width=0.99\textwidth]{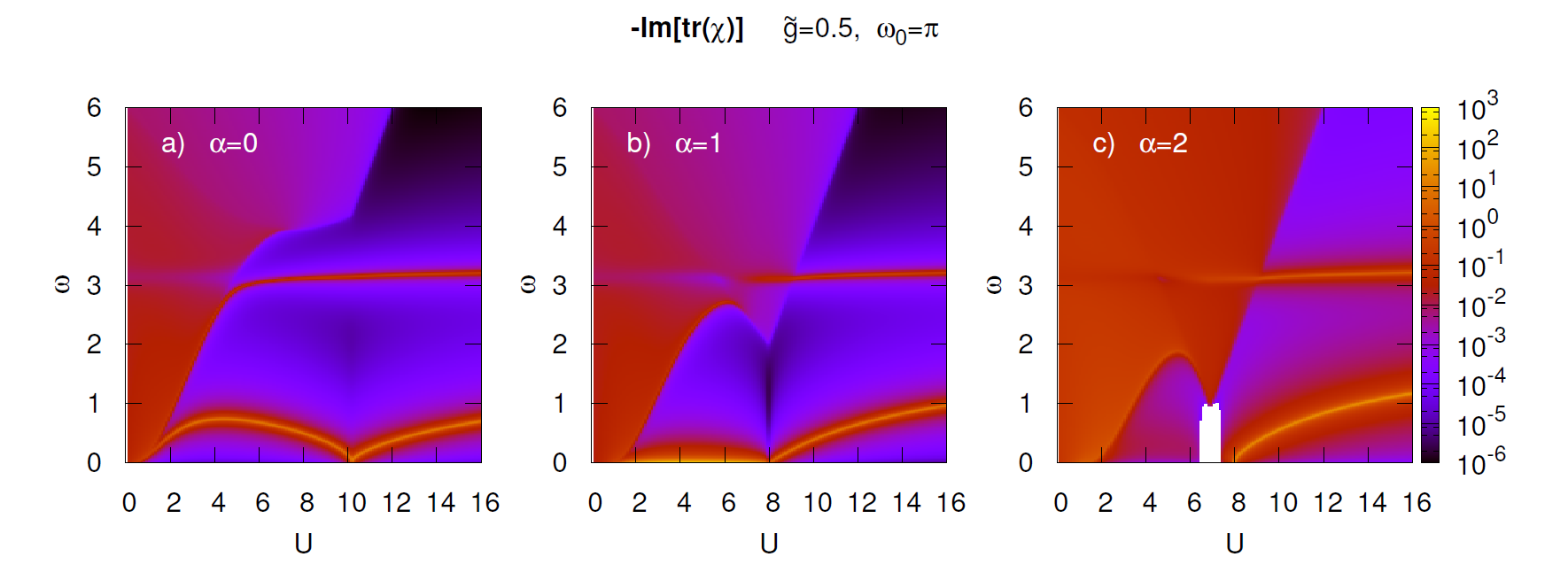}
	\caption{Imaginary part of $-\operatorname{tr}\{\boldsymbol{\chi}\}$ for the one-mode Hamiltonian, for the electron-phonon case ($\alpha=0$, a), the cavity ($\alpha=1$, b), and the overestimated $P^2$ interaction with a real-valued order parameter ($\alpha=2$, c). The white region in panel c) indicates a negative spectral weight (unstable region).}
	\label{fig:sus_phon}
\end{figure*}

\subsubsection{Mean-field solution}

The solution of the one-mode Hamiltonian within mean-field theory is analogous to the treatment of the full model discussed in Sec.~\ref{sec:sc} to \ref{sec:suscept}. We decouple the products $\hat \Pi_0 \hat P$ and $\hat P^2$, and introduce the expectation values $\langle \hat P \rangle =2\phi' L^2$, and $\langle \hat \Pi_0 \rangle  = \sqrt{L^2} \pi_0$. The mean-field oscillator Hamiltonian reads
\begin{equation}
\label{ghjhgfghj001}
	\hat{H}_{mf}^{osc}=\tilde g\sqrt{L^2} \hat \Pi_{0}  2 \phi' +\hat{H}_{free}^{osc},
	\end{equation}
and the electronic mean-field Hamiltonian is defined as in Eq.~\eqref{equ:H_mf_spin}, with the pseudomagnetic field
\begin{equation}
\mathbf{B}_\mathbf{k}=2
\begin{pmatrix}
\tilde g X(\alpha,t)\\
0\\
0
\end{pmatrix}
+2
\begin{pmatrix}
0\\
0\\
\epsilon_\mathbf{k}
\end{pmatrix}
-2U\bm \Phi,
\label{equ:B_k}
\end{equation}
where 
\begin{equation}
X(\alpha, t)=\pi_{0}(t) + \alpha 2\tilde g \phi'.
\end{equation}
We look for static solutions and collective modes by making the ansatz $a(t)=a^{(0)} + \delta a\, e^{-i\omega t}$ for all expectation values, such as $\phi'$ and $\pi_0$. Inserting this into the semiclassical equation of motion 
\begin{align}
\label{someddot}
	\ddot{\pi}_{0}=&-\omega_{0}^2 (\pi_{0} + 2 \tilde g \phi' ),
\end{align}
which can be derived from the oscillator Hamiltonian \eqref{ghjhgfghj001}, yields the static solution 
\begin{align}
\pi_{0}^{(0)}&=
-2\tilde g{\phi'}^{(0)},
\\
\label{x0lambda}
X^{(0)}&=
-(1-\alpha)2\tilde g {\phi'}^{(0)}.
\end{align}
The equilibrium state of the material can be obtained as described in Sec.~\ref{sec:mf-decoupl}, but $-E_0$ must be replaced by $X^{(0)}$, i.e., the equilibrium pseudomagnetic field is  given by
\begin{equation}
\label{bpseudoe0002}
	\mathbf{B}_\mathbf{k}^{(0)}=2
\begin{pmatrix}
-2\tilde g ^2(1-\alpha){\phi'}^{(0)}\\
0\\
0
\end{pmatrix}
+2
\begin{pmatrix}
0\\
0\\
\epsilon_\mathbf{k}
\end{pmatrix}
-2U\bm \Phi^{(0)}.
\end{equation}
Finally turning to the collective modes, the equation of motion \eqref{someddot} gives the response
\begin{align}
\delta \pi_{0}&=
2\tilde g
\frac{\omega_0^2}{\omega^2-\omega_0^2} \delta\phi',
\\
	\delta X&=2\tilde g\Big(\frac{\omega_0^2 }{\omega^2-\omega_0^2}+\alpha\Big)\delta\phi'.
	\label{equ:dX}
\end{align}
With this the induced change in the field $\mathbf{B}$ can be written as
 \begin{equation}
 	\delta\mathbf{B}=(V_U+V_{g}(\alpha))\delta\mathbf{\Phi}
 \end{equation}
and thus takes the same form as Eq.~\eqref{eq:responseB}, where $V_U$ is still defined by Eq.~(\ref{equ:VU}) and
\begin{equation}
	V_{g}(\alpha)=
	4\tilde g^2
	\frac{(1-\alpha)\omega_0^2 +\alpha\omega^2}{\omega^2-\omega_0^2}
	\begin{pmatrix}
	1&0&0\\
	0&0&0\\
	0&0&0
	\end{pmatrix}
	\label{equ:Vph}.
\end{equation}

\subsubsection{Results}

The results for the electron-phonon case ($\alpha=0$) are shown in Fig.~\ref{fig:sus_phon}a, while Fig.~\ref{fig:sus_phon}b corresponds to the cavity case ($\alpha=1$). For $\alpha=0$, the transition (which is marked by the softening of the exciton mode) is shifted to larger interaction strengths $U\approx 10$. Along  with this enhancement of the symmetry-broken phase, the order parameter $|\phi|$ and the single-particle gap are increased, which is reflected in a shift of the boundaries of the  particle-hole continuum. In contrast, the transition for the single-mode cavity ($\alpha=1$) is still at $U=8$ and the magnitude of the order parameter remains unchanged. This behavior can be understood from the mean-field equations: For $\alpha=1$, the static field $X^{(0)}$ [Eq.~\eqref{x0lambda}] vanishes, so that the cavity mode does not affect the equilibrium state of the system. The vanishing of  $X^{(0)}$ can be traced back to the exact cancellation of the dipolar light-matter coupling $\Pi\cdot P$ and the dipolar interaction $P\cdot P$. For $\alpha<1$, the feedback furthermore favors a real order parameter, so that the breaking of the $U(1)$ symmetry due to the coupling to the oscillator directly manifests itself in the static solution. At $\alpha=1$, however, the phase of the order parameter is arbitrary. 

In agreement with the static behavior, the phase mode remains massless for the cavity case ($\alpha=1$), while it becomes massive for $\alpha=0$. The  feedback of the oscillator on the collective modes at low frequencies is determined by the interaction matrix $V_g$ at  $\omega\to0$. Again, one can see that at $\alpha=1$, the contribution from the dipolar interaction $H_{PP}$ and the linear light-matter coupling $H_{EP}$ in Eq.~\eqref{equ:Vph} exactly cancel for $\omega=0$, so that $V_g=0$.

The cancellation of the static feedback of the cavity on the material is fully consistent with classical electrostatics:  In dipolar gauge, the canonical field $-\Pi$ represents the displacement field [c.f.~Eq.~\eqref{doisplsgfe}], so that the feedback $E_0(t)=- \langle\hat P_0 +\hat \Pi_0\rangle/\sqrt{V}$ in Eq.~\eqref{equ:H_sc_matter} and equivalently $X$ in Eq.~\eqref{equ:B_k} represents the electric field. Within classical electrostatics, there is no electric field generated by a material with a homogeneous in-plane polarization, with $\bm \nabla\cdot \bm P=0$. 
\footnote{In contrast, in a geometry different from the
	Fabry-P\'erot 
	cavity, a polarization  generates stray fields. The existence of metallic mirrors will partially quench these external fields, therefore change the energetics of the transition and shift the phase transition.} In contrast, in Coulomb gauge it is more subtle to maintain the correct electrostatics. This can be seen by transforming the Hamiltonian \eqref{sffsddsdsds} back to a representation in which the light-matter coupling enters via the vector potential, represented by the other quadrature $Q_0$ of the cavity mode: Such a transformation is achieved via the unitary transformation 
$\mathcal{W}=e^{i \tilde g/\sqrt{L} \hat P \hat Q_0}$, which shifts $\hat \Pi_0 \to \hat \Pi_0-g/\sqrt{L} \hat P$. The matter Hamiltonian $\hat H_{m}=\hat H_0+\hat H_{int} $ is transformed to $\mathcal{W} \hat H_{m} \mathcal{W}^\dagger \equiv \hat H(Q_0)$. Upon expansion in $Q_0$, 
\begin{align}
&H(Q_0)
=
\hat H_{m}
+
\hat J
\frac{\hat Q_0}{\sqrt{L^2}}
+
\hat K 
\frac{\hat Q_0^2}{L^2}
+
\cdots
\\
&\hat J
=
i[\hat H_{m},\tilde g\hat P],
\,\,\,\,\,\,\,\,
\hat K
=
\frac{1}{2}
[[\hat H_{m}, \tilde g\hat P],\tilde g\hat P],
\end{align}
one arrives at a nonlinear Hamiltonian,\cite{Di-Stefano2019,Andolina2019} with a conventional linear coupling of the vector potential to the current $J \propto dP/dt$, but a nonlinear term $Q_0^2$ which differs from a simple diamagnetic interaction with a coupling of $\hat Q_0^2$ to the density. Hence, in a 
light-matter
Hamiltonian obtained from a straightforward projection of the continuum theory in Coulomb gauge to a restricted set of bands, the classical electrostatic limit is not recovered. 

This result highlights the crucial importance to correctly choose the light-matter Hamiltonian. If the continuum description of the light-matter interaction, $\hat{H}_{EP}= \tfrac12\int d^3\bm r\big[ \hat{\bm{\Pi}}(\bm r) \hat{\bm{ P}}(\bm r)+h.c.\big]/\epsilon(\bm r)$ and $\hat{H}_{PP}=\tfrac12\int d^3\bm r\,\hat{\bm{ P}}(\bm r)^2/\epsilon(\bm r)$, is restricted to a certain subset of states (energy bands in the solid, modes of the electromagnetic field),\cite{Li2020} the balancing of the two parts is only kept if both the polarization $\bm P(\bm r)$ and the field $\bm \Pi(\bm r)$ are represented in the same field modes $\nu$, and if the operator $\bm P_\nu$ in the interaction $H_{PP}$ and $H_{EP}$ [Eqs.~\eqref{equ:H_EP} and \eqref{equ:H_PP}] is projected to the valence band manifold in a consistent manner. For illustrative purposes, we will therefore briefly discuss the consequence of choosing the $P^2$ term too large, i.e., $\alpha>1$. This would arise if the light-matter Hamiltonian was constructed from the continuum by projecting the $\bm P(\bm r)^2$ operator in the polarization energy $\int d^3 \bm r \bm P(\bm r)^2$ to the valence band manifold, instead of taking the square of the projected operator $P(\bm r)$. Fig.~\ref{fig:sus_phon}c shows the spectrum of the single-mode Hamiltonian for $\alpha=2$. Most strikingly, one finds a region of negative spectral weight and a disappearance of the phase mode. This is explained as follows: Because the feedback on $X^{(0)}$ in Eq.~\eqref{bpseudoe0002} has the opposite sign of the one for $\alpha<1$, the stable phase of the order parameter $\phi=|\phi|e^{i\theta}$  is locked to $\theta=\pm \pi/2$ ($\phi'=0$) instead of $\phi''=0$ for $\alpha<1$. Choosing $\theta=0$ (as in Fig.~\ref{fig:sus_phon}c) leads to an unstable solution of the equation with a negative spectral weight. The stable solution at $\theta=\pm\pi/2$ features a massive phase mode, as the solution at $\alpha<0$.

\begin{figure}
	\centerline{
	\includegraphics[width=0.5\textwidth]{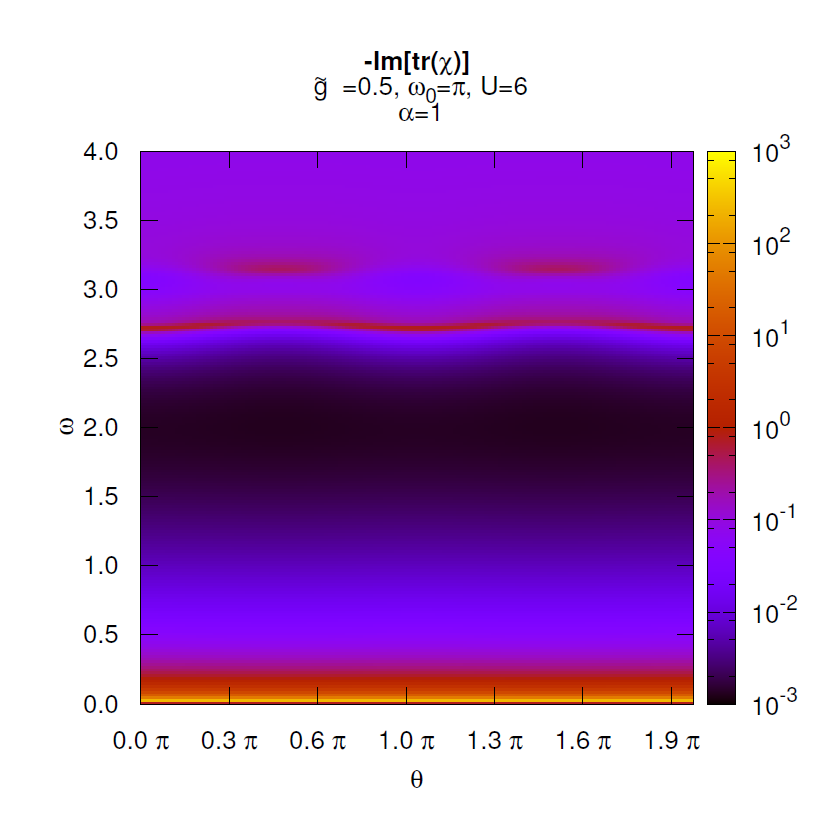}}
	\caption{Imaginary part of $-\operatorname{tr}\{\bm{\chi}\}$  as a function of the frequency $\omega$ and the complex phase $\theta$ of the order parameter $\phi=|\phi|e^{i\theta}$ for a single mode Hamiltonian with $\tilde{g}=0.5$, $\omega_{0}=\pi$, $U=6$ and $\alpha=1$. } 
	\label{fig:somefig}
\end{figure}

\subsection{Breaking of the $U(1)$ symmetry }

The discussion in the previous section has shown how the effect of the dipolar interaction $H_{PP}$ and the light-matter coupling $H_{EP}$ cancel each other. As a consequence, the static mean-field solution is entirely $U(1)$ symmetric, so that the complex phase $\theta$ of the order parameter $\phi=|\phi|e^{i\theta}$ can be chosen arbitrarily. It should be stressed, however, that the coupling to the electromagnetic field nevertheless does reduce the symmetry of the Hamiltonian from $U(1)$ to $Z_2$, even if the $P^2$ term is present ($\alpha=1$). The symmetry is only restored at low frequencies, because the interaction matrix \eqref{equ:Vph} at $\alpha=1$ is proportional to $\omega$. This is illustrated in Fig.~\ref{fig:somefig}, where the spectrum of the single mode Hamiltonian for $\alpha=1$ and $U=6$ is plotted as a function of the phase $\theta$ (in all previous plots, the complex phase was fixed to $\theta=0$). It is clearly  visible that 
the spectral weight for $\theta=0$  is concentrated at lower frequencies than for $\theta=\pi/2$. This may have an interesting consequence: If fluctuations of the order parameter beyond mean-field are included, there is a larger phase space of low energy excitations, which typically implies a larger entropy at finite temperature, and hence a 
stabilization of the phase. Therefore, the finite temperature fluctuations of the order parameter might lead to a breaking of the $U(1)$ symmetry, so that the phase is fixed to $\theta=0,\pi$. 

\section{Conclusion}
\label{sec:conclusion}

In conclusion, we have studied a minimal model for a two-band $U(1)$-symmetric excitonic insulator in a cavity. 
If the electromagnetic field is taken into account, the continuous symmetry of the Hamiltonian is reduced to a discrete $Z_2$ invariance. Nevertheless, the order parameter of the symmetry-broken mean-field state retains an arbitrary $U(1)$ phase, and the phase transition is not affected by the cavity.\cite{Andolina2019}
In dipolar gauge,
the effect can be traced back to a balancing of the linear coupling  $D\cdot P$ between the displacement field and the polarization, and the dipolar self-interaction  $P\cdot P$. The importance of
this mutual elimination has been stressed recently  for atomic systems in strong coupling.\cite{Schaefer2019a}

At nonzero frequencies, light-matter interaction and dipolar self-interaction do not cancel, so that the collective properties at $\omega>0$ depend on the phase of  the order parameter.   A somewhat similar situation is encountered in orbital spin models, where the Hamiltonian is only symmetric under a point group of the lattice, but the manifold of mean-field ground states can have a continuous symmetry. In this case, fluctuations beyond mean-field reflect the lower symmetry and play a crucial role in finding the true phase at finite temperature.\cite{Nussinov2015} It should therefore be interesting to include fluctuations beyond mean-field in the description of the EI in a cavity.

The hybridization between the cavity mode and the material at $\omega>0$ can have other interesting consequences. For example, it can push the amplitude mode of the condensate out of the particle-hole continuum and make it long-lived.  
Moreover, it is questionable if the U(1)-symmetric model for the EI captures the behaviour of real materials. Rather, one would expect that phonons as well as inter-band hybridizations already reduce the continuous U(1) invariance to some point group symmetry of the lattice. 
Because collective modes at nonzero frequency are affected by the light-matter interaction
it may be possible to control the behaviour of such a material inside a cavity through nonlinear driving. 
Also for these situations, our results show that it is  important to properly include the nonlinearities ($A^2$ of $P^2$) arising from the light-matter coupling.

\acknowledgments
We acknowledge discussions with D.~Jaksch, G.~Mazza, A.~Georges, A.~Millis, D.~Golez, C. Sch\"afer, and J.~Li.  This work was supported by the ERC starting grant No. 716648.

%\bibliographystyle{apsrev4-1}
%\bibliography{apssamp01}

%merlin.mbs apsrev4-1.bst 2010-07-25 4.21a (PWD, AO, DPC) hacked
%Control: key (0)
%Control: author (72) initials jnrlst
%Control: editor formatted (1) identically to author
%Control: production of article title (-1) disabled
%Control: page (0) single
%Control: year (1) truncated
%Control: production of eprint (0) enabled
%

\end{document}